# A Suggestion for $^{10}$B Imaging During Boron Neutron Capture Therapy

M. Cortesi

Department of Physics of the University and INFN, Milan, Italy.

**ABSTRACT**

Selective accumulation of $^{10}$B compound in tumour tissue is a fundamental condition for the achievement of BNCT (Boron Neutron Capture Therapy), since the effectiveness of therapy irradiation derives just from neutron capture reaction of $^{10}$B. Hence, the determination of the $^{10}$B concentration ratio, between tumour and healthy tissue, and a control of this ratio, during the therapy, are essential to optimise the effectiveness of the BNCT, which it is known to be based on the selective uptake of $^{10}$B compound. In this work, experimental methods are proposed and evaluated for the determination *in vivo* of $^{10}$B compound in biological samples, in particular based on neutron radiography and gamma-ray spectroscopy by telescopic system. Measures and Monte Carlo calculations have been performed to investigate the possibility of executing imaging of the $^{10}$B distribution, both by radiography with thermal neutrons, using $^6$LiF/ZnS:Ag scintillator screen and a CCD camera, and by spectroscopy, based on the revelation of gamma-ray reaction products from $^{10}$B and the $^1$H. A rebuilding algorithm has been implemented. The present study has been done for the standard case of $^{10}$B uptake, as well as for proposed case in which, to the same carrier, is also synthesized $^{157}$Gd, in the amount of is used like a contrast agent in NMRI.

## Introduction

At present, any routine technique does not exist for the determination *in-vivo* of $^{10}$B distribution in tissue in real time during the therapy. Direct methods are in study in various research laboratory to obtain a qualitative pre-treatment imaging, for instance based on Positron Emission Tomography (PET) or Nuclear Magnetic Resonance Imaging (NMRI), or indirect methods based on analysis of a patient blood sample collected before neutron treatment.

However, pharmacokinetic investigations have clearly demonstrated that wide variability exists in the relation to uptake of $^{10}$B compound between tumour tissue and blood. Moreover, the time dependency of the relative concentration exhibits large discrepancy for different patients, also in case of patients with the same pathologies. In fact, the delivering of $^{10}$B depends upon vary factors, and includes in particular way the perfusion, the permeability of vessel in the tumour tissue, and the cellular differentiation.

Lately, some of the most important existing operating units are beginning a study on the applicability of the gamma-ray spectroscopy, such as technique for the on-line control of $^{10}$B compound in tissue during BNCT irradiation.

## Neutron Radiography

The neutron radiographical image is obtained with a direct digital system: the neutron flux activates a screen converter, placed behind the object under investigation. The screen converter emits prompt radiation that impresses a sensitive film. In the case of a scintillator screen it is possible to use a CDD-camera, which detects the light output of the scintillator screen. Neutrons are detected by a typical chain, shown in Fig.1: [$^6$LiF/ZnS:Ag screen] → [Image intensifier] → [CCD-camera, ST6 camera of Santa Barbara Instruments Group, with a CCD sensor TC241 of Texas Instruments][1]. The CCD sensor is presently used has a sensible area of 2.64 mm x 2.64 mm (with a 169x191 pixel matrix). The neutron radiography system has been developed at the thermal column of the 1 MW TRIGA Mark II reactor, running at ENEA Casaccia Research Centre.

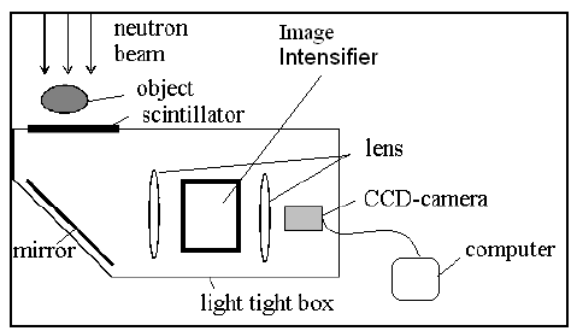

Fig.1: Set-up of a CCD camera detector system for neutron radiography.

Monte Carlo simulations have been performed, using MNCP4c3 code, in order to verify the feasibility of the neutron radiography like a direct method for $^{10}$B and $^{157}$Gd imaging in biological sample. Assuming that the MCNP-model-phantom contains only two regions with different $^{10}$B or $^{157}$Gd concentrations, various situation have been studied: in the hypothesis that in the tumoural compartment there are neither $^{10}$B nor $^{157}$Gd, in the hypothesis that in the tumoural compartment there is the $^{10}$B (40 ppm), in the hypothesis that there is $^{157}$Gd (100 ppm) and, lastly, in the hypothesis that there is the presence of both $^{10}$B and $^{157}$Gd, in the above cited concentrations. The MCNP results are shown in Fig.3, where for each configuration, the result obtained by subtracting the profile in the phantom with tumour from that in the phantom without tumour, is reported.

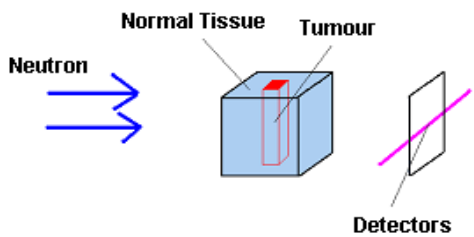

Fig. 2: Set-up of Monte Carlo simulation

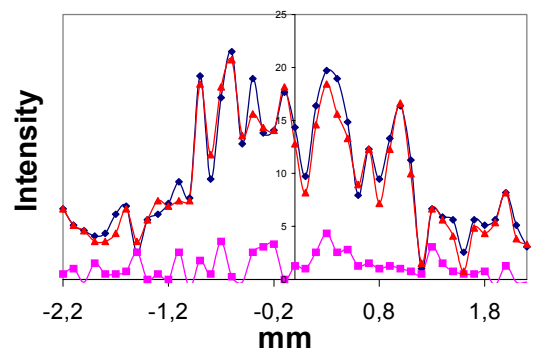

Fig. 3: Result of MCNP simulation
($^{10}$B, $^{157}$Gd, $^{10}$B+$^{157}$Gd)

Neutron radiography scansions are obtained simulating irradiations of 50 seconds with a neutron flux density of $10^7$ cm$^{-2}$ s$^{-1}$. The MCNP model of detector for radiographical scansion has assumed the same characteristics of sensibility and efficiency of that utilised in this work. According with MNCP results, it is clear that the presence of $^{157}$Gd, both with (line blue) or without (line red) $^{10}$B, is detectable. On the opposite, the presence in the tumour of only $^{10}$B is hardly perceivable (line purple).

A MaCoStudio software has been developed for the digital elaboration of neutron radiographic images, in order to correct the space and temporal disomogenity of the thermal flux[2]. MaCoStudio is able to compare different neutron radiographies and to put in evidence the presence of elements with high thermal neutron cross section, like $^{10}$B and $^{157}$Gd.

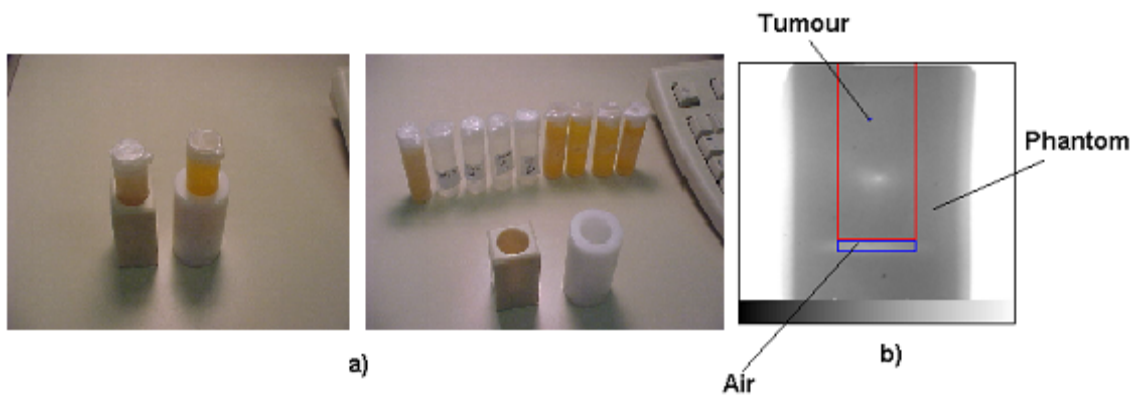

Fig. 4: a) Test objects for neutron radiography system performance evalutation
b) Sample neutron radiography

In order to evaluate the possibility of performing the neutron radiography of biological tissue, some tests with hydrogenous samples have been carried out. The specimens (shown in Fig. 4) were made of polyethylene, containing a hole that can be filled with a solution containing various concentrations of $^{10}$B and $^{157}$Gd.

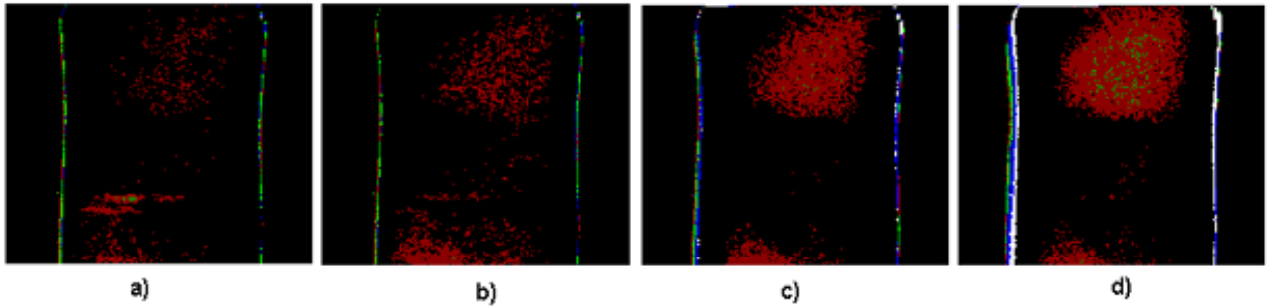

Fig. 5: Neutron Radiography; a) 35 ppm of $^{10}$B, b) 1000 ppm of $^{10}$B, c) 100 ppm of $^{157}$Gd, d) 1000 ppm of $^{157}$Gd

The images, presented in Fig.5, show the reconstruction of $^{10}$B and $^{157}$Gd concentrations in the investigated samples, performed by the digital elaboration with MaCoStudio software of the images obtained with the radiographical system.

**Prompt Gamma-Ray Spectroscopy**

As known, following the thermal neutron capture reactions, prompt γ-rays are emitted by $^{10}$B; from the rate of γ-rays emitted in a small specific region, it's possible to calculate the $^{10}$B concentrations with an easy formalism[3][4][5][6].

Monte Carlo simulations have been performed, using MNCP4c3 code, in order to verify the feasibility of the prompt γ-ray spectroscopy like a direct method for $^{10}$B concentration measurements in biological samples. A water phantom has been considered, in which a tumoural compartment with amounts of $^{10}$B concentration are present. By a simulated telescope system, just a small region is scanned. Assuming that the $^{10}$B concentrations is homogeneously distributed in the tumour and in surrounding tissue, only two regions have to be considered: healthy tissue and tumoural compartment.

Monte Carlo calculations have shown that $^{10}$B detection and concentration reconstruction, in correspondence of tumoural compartment, are achieved with an acceptable standard deviation. However, high resolution spectroscopy and a short measurement time are needed; both conditions are realized by proper optimisation of the collimator system and of the radiation shield for detector, owing to the high background[5].

**Discussion**

It has been shown that neutron radiography can be fruitfully applied like direct method for determination of $^{10}$B and $^{157}$Gd accumulations in small hydrogenous samples. In particular, the technique obtains better results with $^{157}$Gd, even if in a lower concentration like that possibly used for BNCT, in addiction to $^{10}$B. In fact, in the hypothesis that $^{10}$B and $^{157}$Gd could be synthesized to the same carrier, the qualitative determination of the $^{157}$Gd spatial distribution is representative of $^{10}$B distribution.

The γ-ray spectroscopy technique may allow determining the $^{10}$B concentrations by detecting the photons emitted from $^{10}$B reactions, using high efficiency detectors and fine telescope systems.

## Perspectives

Presently, at ENEA Casaccia Research Centre it is in phase of development a further neutron-radiography-acquisition system that will allow to improve the resolution of the neutron radiographical image and to wide the field of view. Moreover, in the near future, the design and implementation of a new near parallel collimator is planned. This would greatly enhance the testing capabilities of the entire neutron radiography system, now limited to small size objects.

Besides $^{10}$B concentration measurements, *in vivo* dosimetry will become a future goal for verification of treatment planning, and the γ-ray spectroscopy may provide a tool for this purpose[7]. As said before, the feasibility of pratical γ-ray spectroscopy depends on further optimization of collimator and of radiation shield[5].

## Acknowledgements

The work was partially supported by INFN (Italy). The author is grateful to working group of ENEA Casaccia Research Centre, and in particular Dr. R. Rosa which is responsible for developing the employed techniques for neutron radiography, Dr. N. Burgio and Dr. A. Grossi. Lastly, the author would like to thank to all the staff of the FriXy and TLD laboratory (Milan University, Italy), in particular Prof. G. Gambarini.